\documentstyle{article}

\begin{document}

\newcommand{\nc}[2]{\newcommand{#1}{#2}}
\newcommand{\ncx}[3]{\newcommand{#1}[#2]{#3}}
\ncx{\pr}{1}{#1^{\prime}}
\nc{\nl}{\newline}
\nc{\np}{\newpage}
\nc{\nit}{\noindent}
\nc{\be}{\begin{equation}}
\nc{\ee}{\end{equation}}
\nc{\ba}{\begin{array}}
\nc{\ea}{\end{array}}
\nc{\bea}{\begin{eqnarray}}
\nc{\eea}{\end{eqnarray}}
\nc{\nb}{\nonumber}
\nc{\dsp}{\displaystyle}
\nc{\bit}{\bibitem}
\nc{\ct}{\cite}
\ncx{\dd}{2}{\frac{\partial #1}{\partial #2}}
\nc{\pl}{\partial}
\nc{\dg}{\dagger}
\nc{\cL}{{\cal L}}
\nc{\ag}{\alpha}
\nc{\bg}{\beta}
\nc{\gam}{\gamma}
\nc{\Gam}{\Gamma}
\nc{\bgm}{\bar{\gam}}
\nc{\del}{\delta}
\nc{\Del}{\Delta}
\nc{\eps}{\epsilon}
\nc{\ve}{\varepsilon}
\nc{\th}{\theta}
\nc{\vt}{\vartheta}
\nc{\kg}{\kappa}
\nc{\lb}{\lambda}
\nc{\ps}{\psi}
\nc{\Ps}{\Psi}
\nc{\sg}{\sigma}
\nc{\spr}{\pr{\sg}}
\nc{\Sg}{\Sigma}
\nc{\rg}{\rho}
\nc{\fg}{\phi}
\nc{\Fg}{\Phi}
\nc{\vf}{\varphi}
\nc{\og}{\omega}
\nc{\Og}{\Omega}
\nc{\Kq}{\mbox{$K(\vec{q},t|\pr{\vec{q}\,},\pr{t})$ }}
\nc{\Kp}{\mbox{$K(\vec{q},t|\pr{\vec{p}\,},\pr{t})$ }}
\nc{\vq}{\mbox{$\vec{q}$}}
\nc{\qp}{\mbox{$\pr{\vec{q}\,}$}}
\nc{\vp}{\mbox{$\vec{p}$}}
\nc{\va}{\mbox{$\vec{a}$}}
\nc{\vb}{\mbox{$\vec{b}$}}
\nc{\Ztwo}{\mbox{\sf Z}_{2}}
\nc{\Tr}{\mbox{Tr}}
\nc{\lh}{\left(}
\nc{\rh}{\right)}
\nc{\cB}{\mbox{$^{\ast}\Og$ }}
\nc{\nil}{\emptyset}
\nc{\bor}{\overline}
\renewcommand{\thepage}{}

\pagestyle{empty}

\begin{flushright}
                                                     NIKHEF-H/92-08
\end{flushright}
\vspace{2ex}

\begin{center}

{\LARGE \bf Symmetries and Motions in Manifolds\footnote{Lectures at the 28th
        Karpacz Winterschool of Theoretical Physics (Poland, 1992), by
      J.W.\ van Holten} }\\
\vspace{3ex}
{\large J.W.\ van Holten\footnote{Research supported by the Stichting FOM} and
        R.H.\ Rietdijk\footnote{Address after oct.\ 1, 1992:
        Dept.\ of Mathematics, Science Labs., Univ.\ of Durham U.K.} }  \\
        \vspace{1ex}
        NIKHEF-H, P.O.\ Box 41882, 1009 DB Amsterdam (NL)
\end{center}
\vspace{4ex}

{\small
 \begin{center}
   {\bf Abstract} \\ \vspace{1ex}

  \begin{minipage}{30em}
  In these lectures the relations between symmetries, Lie algebras, Killing
  vectors and Noether's theorem are reviewed. A generalisation of the basic
ideas
  to include velocity-dependend co-ordinate transformations naturally leads to
  the concept of Killing tensors. Via their Poisson brackets these tensors
  generate an {\em a priori} infinite-dimensional Lie algebra. The nature of
such
  infinite algebras is clarified using the example of flat space-time. Next the
  formalism is extended to spinning space, which in addition to the standard
real
  co-ordinates is parametrized also by Grassmann-valued vector variables. The
  equations for extremal trajectories (`geodesics') of these spaces describe
the
  pseudo-classical mechanics of a Dirac fermion. We apply the formalism to
solve
  for the motion of a pseudo-classical electron in Schwarzschild space-time.
  \end{minipage}
 \end{center}
 }
\np

\pagestyle{plain}
\renewcommand{\thepage}{\arabic{page}}
\setcounter{page}{1}

\section{Motions of Scalar Points in Curved Space-Time}

\subsection{Introduction}

In the following we connect a number of old and venerable topics related
to symmetries and conservation laws, such as Lie algebras and Noether's
theorem,
with differential geometric structures like Lie derivatives and Killing
vectors.
Although most of the basic ideas are well-known\footnote{See for example
refs.\ct{MTW}-\ct{KSmCH}.}, we present extensions and generalisations of
interest in the description of certain physical systems; in particular we apply
our methods to study the motion of spinning particles in a curved space-time.

According to Einstein's equivalence principle the world line of a massive
scalar
point particle in curved space-time is a time-like geodesic, described by the
equation

\be
\frac{D^{2} x^{\mu}}{D \tau^{2}}\, = \ddot{x}^{\mu}\, +\,
  \Gam_{\lb \nu}^{\:\:\:\mu}\, \dot{x}^{\lb} \dot{x}^{\nu} = 0,
\label{1.1}
\ee

\nit
with overdots denoting proper time derivatives. Its time-like nature is
expressed by the condition

\be
\left( \frac{ds}{d\tau} \right)^{2} = g_{\mu\nu}\, \frac{d x^{\mu}}{d \tau}
\frac{d x^{\nu}}{d \tau} = - c^{2} < 0,
\label{1.2}
\ee

\nit
where the universal constant $c$ (the light velocity) is a real number. In the
following we usually take $c=1$ and consider particles of unit mass, but
occasionally we re-instate the explicit mass dependence when this is physically
relevant.

Note, that eq.(\ref{1.1}) implies that the acceleration due to gravity is
quadratic in the four-velocity, whereas according to the Lorentz force law the
acceleration of a particle subject to electro-magnetic forces is linear in the
four-velocity. For particles coupled to force fields of higher spin one expects
the acceleration to depend on higher powers of the four-velocity \ct{dWF}.

The geodesic law of motion (\ref{1.1}) can be derived from an action principle.
The simplest form of the action is

\be
S = \int_{1}^{2} d\tau\, \frac{1}{2}\, g_{\mu\nu}(x) \dot{x}^{\mu}
    \dot{x}^{\nu},
\label{1.3}
\ee

\nit
the stationary points of which are precisely given by eq.(\ref{1.1}). Indeed,
the general variation of $S$ reads

\be
\del S = \int_{1}^{2} d\tau\, \left\{ - \del x^{\mu}\, g_{\mu\nu}
         \frac{D^{2}x^{\nu}}{D \tau^{2}}\, +\, \frac{d}{d\tau}\left(
         \del x^{\mu} p_{\mu} \right) \right\},
\label{1.4}
\ee

\nit
where $p_{\mu} = g_{\mu\nu} \dot{x}^{\nu}$ is the canonical momentum. Thus
$\del S$ vanishes for any arbitrary variation of $x^{\mu}$ with fixed end
points
if and only if the equations of motion (\ref{1.1}) are satisfied.

\subsection{Symmetries and Noether's Theorem}

In regard to eq.(\ref{1.4}) we can now ask, whether there exist variations
$\del x^{\mu}$ for which $\del S = 0$ modulo boundary terms even when the
equations of motion are {\em not} satisfied:

\be
\del S = \int_{1}^{2} d\tau\, \frac{d}{d\tau} \left( \del x^{\mu} p_{\mu} -
         {\cal J}(x,\dot{x}) \right).
\label{2.1}
\ee

\nit
Here we have already split off the total derivative coming from partial
integrations in the derivation of expression (\ref{1.4}). The quantity
${\cal J}(x,\dot{x})$ is obtained from variations $\del x^{\mu}$ of the type

\be
\del x^{\mu} = {\cal R}^{\mu}(x,\dot{x}) = R^{\mu}(x)\, +\, \dot{x}^{\nu}
               K_{\nu}^{\:\:\mu}(x)\, +\, \frac{1}{2} \dot{x}^{\nu}
               \dot{x}^{\lb} L_{\nu\lb}^{\:\:\:\:\mu}(x)\, + ...
\label{2.2}
\ee

\nit
We restrict ourselves to variations which depend on the first derivative
$\dot{x}^{\mu}$ only, because the second and higher derivatives can always be
rewritten in terms of these modulo the equations of motion (\ref{1.1}).
Comparing eqs.(\ref{1.4}) and (\ref{2.1}) one immediately finds

\be
\frac{d {\cal J}}{d\tau}\, =\, {\cal R}^{\mu}\, g_{\mu\nu}\,
      \frac{D^{2}x^{\nu}}{D \tau^{2}} \approx 0,
\label{2.3}
\ee

\nit
where the last equality holds only upon using the equations of motion. Hence
for physical motions the quantities ${\cal J}(x,\dot{x})$ are conserved. This
is Noether's theorem.

\subsection{Generalised Killing Equations}

Assuming that ${\cal J}(x,\dot{x})$ can be expanded in the four-velocity as

\be
{\cal J}(x,\dot{x}) = J^{(0)}(x) + \dot{x}^{\mu} J^{(1)}_{\mu}(x) +
    \frac{1}{2}\, \dot{x}^{\mu} \dot{x}^{\nu} J^{(2)}_{\mu\nu}(x) +
    \frac{1}{3!}\, \dot{x}^{\mu} \dot{x}^{\nu} \dot{x}^{\lb}
    J^{(3)}_{\mu\nu\lb}(x) + ...
\label{3.1}
\ee

\nit
we can compare terms with equal powers of the acceleration and velocity on the
left- and right-hand side of eq.(\ref{2.3}), using the Ansatz (\ref{2.2}) for
$\del x^{\mu}$. This leads first of all to an identification of the
co-efficients in the expansions (\ref{2.2}) and (\ref{3.1}):

\be
\ba{lll}
J^{(1)}_{\mu}(x) & = & R_{\mu}(x), \\
  &  &  \\
J^{(2)}_{\mu\nu}(x) & = & K_{\mu\nu}(x), \\
  &  &  \\
J^{(3)}_{\mu\nu\lb}(x) & = & L_{\mu\nu\lb}(x), \hspace{2cm} etc. \\
\ea
\label{3.2}
\ee

\nit
indicating that all covariant tensors on the right hand side of eq.(\ref{2.2})
should be taken to be completely symmetric. Secondly, the comparison shows that
the following differential equations have to be satisfied

\be
J^{(n)}_{\left( \mu_{1}...\mu_{n};\mu_{n+1} \right)} = 0,
\label{3.3}
\ee

\nit
in which the parentheses denote full symmetrisation over all indices enclosed,
with total weight one. Eqs.(\ref{3.3}) constitute a straightforward
generalisation of the Killing equation for the isometries of differentiable
manifolds. Explicitly

\begin{eqnarray}
J^{(0)}_{,\mu} & = & 0,
\label{3.4a} \\
  &  & \nonumber \\
R_{\left( \mu;\nu \right)} & = & 0,
\label{3.4b} \\
  &  & \nonumber \\
K_{\left( \mu\nu;\lb \right)} & = & 0, \hspace{2cm} etc.
\label{3.4c}
\end{eqnarray}

\nit
The first equation (\ref{3.4a}) implies that $J^{(0)}$ is an irrelevant
constant
which we ignore from now on. The second equation (\ref{3.4b}) is the standard
equation for Killing vectors, whilst (\ref{3.4c}) and its higher-rank
counterparts constitute tensorial generalisations of this equation. Therefore
one refers to $K_{\mu\nu}$ and higher-rank tensors satisfying eq.(\ref{3.3}) as
Killing tensors.

\subsection{Canonical Analysis}

In terms of phase-space variables $(x^{\mu},p_{\mu})$ the conserved quantities
of motion read

\be
{\cal J}(x,p) = p_{\mu} R^{\mu}(x)\, +\, \frac{1}{2}\, p_{\mu}p_{\nu}
   K^{\mu\nu}(x)\, +\, \frac{1}{3!}\, p_{\mu} p_{\nu} p_{\lb} L^{\mu\nu\lb}(x)
   + ...
\label{4.1}
\ee

\nit
Suppose that there exist $n$ independent Killing vectors $R_{a}^{\mu}(x)$,
$a = 1,...,n$. Then a general Killing vector is a linear combination

\be
R^{\mu}[\xi] = \xi^{a} R^{\mu}_{a},
\label{4.2}
\ee

\nit
where the $\xi^{a}$ constitute a set of $n$ linearly independent parameters.
A similar remark holds for Killing tensors. Hence a particular ${\cal J}$,
eq.(\ref{4.1}), is in general specified by the values of these parameters
$\xi^{A} = (\xi^{a},...)$.

Introducing the fundamental Poisson brackets

\be
\left\{ x^{\mu}, p_{\nu} \right\} = \del_{\nu}^{\mu},
\label{4.3}
\ee

\nit
we can compute the Poisson bracket of two conserved Noether charges
${\cal J}(1) \equiv {\cal J}[\xi^{A}_{1}]$ and ${\cal J}(2) \equiv
{\cal J}[\xi^{A}_{2}]$. The result is

\be
\ba{lll}
\left\{ {\cal J}(1), {\cal J}(2) \right\} & = & \dsp{
    p_{\mu} \left( R^{\mu}(1)_{;\lb} R^{\lb}(2)\, -\, R^{\mu}(2)_{;\lb}
    R^{\lb}(1) \right) } \\
  & & \\
  & + & \dsp{ \frac{1}{2}\, p_{\mu} p_{\nu} \left( K^{\mu\nu}(1)_{;\lb}
         R^{\lb}(2)\, -\, K^{\mu\nu}(2)_{;\lb} R^{\lb}(1)\, \right. } \\
  & & \\
  & & \dsp{ \left. +\, 2 R^{\mu}(1)_{;\lb} K^{\nu\lb}(2)\, -\,
      2 R^{\mu}(2)_{;\lb} K^{\nu\lb}(1) \right) } \\
  & & \\
  & + & \dsp{ \frac{1}{3!}\, p_{\mu} p_{\nu} p_{\lb} \left(
        L^{\mu\nu\lb}(1)_{;\kg} R^{\kg}(2)\, -\, L^{\mu\nu\lb}(2)_{;\kg}
        R^{\kg}(1)\,  \right. } \\
  & & \\
  & & \dsp{ -\, 3 K^{\kg \left( \mu\nu \right. }(1) R^{\left.
      \lb \right) }(2)_{;\kg}\, +\, 3 K^{\kg \left( \mu\nu \right.}(2)
      R^{\left. \lb \right)}(1)_{;\kg}\, } \\
  & & \\
  & & \dsp{ \left.  +\, 3 K^{\left( \mu\nu \right.}_{;\kg}(1)
      K^{\left. \lb \right) \kg}(2)\, -\, 3 K^{\left( \mu\nu \right.}_{;\kg}(2)
      K^{\left. \lb \right) \kg}(1) \right)\, } \\
  & & \\
  & + & ... \\
\ea
\label{4.4}
\ee

\nit
The left-hand side being conserved for physical motions, the right-hand side
must again be of the form (\ref{4.1}), modulo equations of motion. We now
restrict ourselves to off-shell closed algebras, i.e.\ the case in which the
Poisson bracket (\ref{4.4}) itself takes the form of a Noether charge without
explicit use of the equations of motion. Then

\be
\left\{ {\cal J}(1), {\cal J}(2) \right\} = {\cal J}(3),
\label{4.5}
\ee

\nit
where the parameters $\xi_{3}$ are bilinear combinations of $\xi_{1}$ and
$\xi_{2}$:

\be
\xi^{A}_{3} = f^{BCA}\, \xi^{B}_{2} \xi^{C}_{1}.
\label{4.6}
\ee

\nit
{}From eqs.(\ref{4.4})-(\ref{4.6}) it now follows, that

\begin{eqnarray}
\left[ \cL_{R^{a}}(R) \right]^{b \mu} & \equiv &
       R^{b \mu}_{\:\:\:;\nu} R^{a \nu}\, -\, R^{a \mu}_{\:\:\:;\nu} R^{b \nu}
       = f^{abc} R^{c \mu},
\label{4.7a} \\
  &  & \nonumber \\
\left[ \cL_{R^{a}}(K) \right]^{i\,\mu\nu} & \equiv &
       K^{i\,\mu\nu}_{\:\:\:\:\:;\lb} R^{a\lb}\, - \, 2 R^{a \left( \mu
       \right.}_{\:\:\:\:;\lb} K^{i\, \left. \nu \right) \lb}
       = t^{aij} K^{j\,\mu\nu},
\label{4.7b} \\
 &  &  \nonumber \\
\left[ \cL_{R^{a}}(L) \right]^{r\,\mu\nu\lb} & \equiv &
       L^{r\mu\nu\lb}_{\:\:\:\:\:\:\:\:;\kg} R^{a\kg}\, -\,
       3 R^{a\left( \mu \right.}_{\:\:\:\:;\kg} L^{r \left. \nu\lb\right) \kg}
       =  g^{ars} L^{s\,\mu\nu\lb}, \hspace{2em} etc.
\label{4.7c}
\end{eqnarray}

\nit
Here the symbol $\cL$ has been introduced to denote Lie derivatives. In
addition,
we find relations involving higher-rank Killing tensors of the type

\be
\left[ \cL_{K^{j}}(K) \right]^{i\,\mu\nu\lb}\, \equiv\,
       K^{i\left( \mu\nu \right.}_{\:\:\:\:\:\:;\kg} K^{j \left. \lb \right)
\kg}
       \, -\, K^{j\left( \mu\nu \right.}_{\:\:\:\:\:\:;\kg} K^{i \left. \lb
       \right) \kg}\, =\, c^{ijr} L^{r\,\mu\nu\lb},
\label{4.8}
\ee

\nit
and its further generalisations. Thus the notion of Lie derivative is extended
to higher-rank tensors.

Observe, that eq.(\ref{4.7a}) implies that the Killing vectors define an
$n$-dimensional Lie algebra. Equations like (\ref{4.7b}), (\ref{4.7c}) then
assert that the Killing tensors transform in linear representations of this
Lie algebra defined by the structure constants $(t^{aij}, g^{ars},...)$.
Indeed,
the Jacobi identities guarantee that these structure constants realize the
Lie algebra via their matrix commutators:

\be
\ba{lll}
\left[ t^{a}, t^{b} \right]^{ij} & = & - f^{abc}\, t^{cij}, \\
  &  & \\
\left[ g^{a}, g^{b} \right]^{rs} & = & - f^{abc}\, g^{crs}, \hspace{4em} etc.
\\
\ea
\label{4.9}
\ee

\nit
{}From eq.(\ref{4.8}) it folllows, that Killing tensors of rank $n$ and $m$
generate Killing tensors of rank $(n+m-1)$ via their tensorial Lie derivatives.
In this way one obtains in principle an infinite-dimensional algebra of
conserved quantities, unless the left-hand side of eq.(\ref{4.8}) vanishes
identically, as might happen in special cases.

\subsection{A Universal Solution}

The generalised Killing equations (\ref{3.3}) admit one solution which exists
for any arbitrary metric $g_{\mu\nu}$. This solution is generated by the metric
itself:

\be
K_{\mu\nu}(x) = g_{\mu\nu}(x).
\label{5.1}
\ee

\nit
It satisfies eq.(\ref{3.3}) identically by virtue of the metric postulate
$g_{\mu\nu;\lb} = 0$. The constant of motion constructed from this Killing
tensor is the world-line Hamiltonian

\be
H(x,p) = \frac{1}{2}\, g^{\mu\nu}(x)\, p_{\mu} p_{\nu},\
\label{5.2}
\ee

\nit
which is the generator of proper-time translations. From this solution we can
construct a whole tower of higher order Killing tensors by taking completely
symmetrised products of $n$ metric tensors. The corresponding conserved
quantities are given by the $n$-th power of the Hamiltonian (\ref{5.2}).
Clearly, all these Noether charges commute among themselves, and no other
solutions are generated by their Poisson brackets. The existence of any further
solutions to the generalised Killing equations depends on the specific choice
of $g_{\mu\nu}(x)$.

Note, that the method of constructing higher-rank Killing tensors out of
products of lower ones works quite generally. Indeed, from any two Killing
tensors $J^{(n)}_{\mu_{1}...\mu_{n}}$ and $J^{(m)}_{\mu_{1}...\mu_{m}}$
one can construct a new Killing tensor

\be
J^{(n+m)}_{\mu_{1}...\mu_{n+m}} = J^{(n)}_{\left( \mu_{1}...\mu_{n} \right.}
           J^{(m)}_{\left. \mu_{1}...\mu_{m} \right)}.
\label{5.3}
\ee

\nit
This satisfies the generalised Killing equation (\ref{3.3}) because of Leibniz'
rule.

\subsection{Example: Flat Space}

In order to illustrate the general formalism presented above, we consider the
example of flat space, for which all conserved quantities of motion can be
constructed explicitly. For

\be
g_{\mu\nu}(x) = \del_{\mu\nu},
\label{6.1}
\ee

\nit
the Killing equation

\be
R_{\mu,\nu} + R_{\nu,\mu} = 0
\label{6.2}
\ee

\nit
has the general solution

\be
R_{\mu}[a,\og] = a_{\mu} + \og_{\mu\nu} x^{\nu}, \hspace{1.5cm}
   \og_{\mu\nu} = - \og_{\nu\mu}.
\label{6.3}
\ee

\nit
Here $a_{\mu}$ and $\og_{\mu\nu}$ are constant parameters labeling the various
independent Killing vectors (ten in four dimensions). The constants of motion
corresponding to these Killing vectors are:

\be
J^{(1)}[a,\og] = a_{\mu} p^{\mu} + \frac{1}{2}\, \og_{\mu\nu} M^{\mu\nu},
\label{6.4}
\ee

\nit
with $M^{\mu\nu} = x^{\mu} p^{\nu} - x^{\nu} p^{\mu}$. Clearly, the first term
in eq.(\ref{6.4}) generates a translation, whilst the second one generates
Lorentz transformations.

At the level of second-rank tensors one finds ---in addition to symmetrised
products of Killing vectors--- the universal solution $\del_{\mu\nu}$ plus
one new solution:

\be
K_{\mu\nu}[\ag,\bg] = \ag\, \del_{\mu\nu}\, +\, \bg \left( \del_{\mu\nu} x^{2}
   - x_{\mu} x_{\nu} \right).
\label{6.5}
\ee

\nit
These solutions correspond to the quadratic Casimir invariants of the
Poincar\'{e} algebra:

\be
J^{(2)}[\ag,\bg] = \ag\, p_{\mu}^{2}\, +\, \frac{1}{2}\, \bg M_{\mu\nu}^{2}.
\label{6.6}
\ee

\nit
Since for scalar particles there exist no other independent higher order
Casimir invariants of the Poincar\'{e} algebra, all other Killing tensors can
now be expressed as symmetrised products of the Killing vectors (\ref{6.3}) and
the second-rank Killing tensors (\ref{6.5}). For example, the Killing tensors
of
rank 3 which can be constructed are of the form

\be
\ba{lll}
L_{\mu\nu\lb}[a^{(1)},\og^{(1)}] & = & \del_{\left( \mu\nu \right.} \left(
   a^{(1)}_{\left. \lb \right)} + \og^{(1)}_{\left. \lb \right) \kg} x^{\kg}
   \right) \\
  &  &  \\
L_{\mu\nu\lb}[a^{(2)},\og^{(2)}] & = & \left( \del_{\left( \mu\nu \right.}
x^{2}
   - x_{\left( \mu \right.} x_{\nu} \right) \left( a^{(2)}_{\left. \lb \right)}
   + \og^{(2)}_{\left. \lb \right) \kg} x^{\kg} \right). \\
\ea
\label{6.7}
\ee

\nit
To these tensors correspond the Noether charges

\be
J^{(3)}[a^{(i)},\og^{(i)}] = p^{2} \left( a^{(1)} \cdot p + \frac{1}{2}\,
   \og^{(1)} \cdot M \right) + \frac{1}{2}\, M^{2} \left( a^{(2)} \cdot p +
   \og^{(2)} \cdot M \right).
\label{6.8}
\ee

\nit
Clearly, the constants of motion $(p_{\mu}, M_{\mu\nu})$ form the building
blocks for the construction of the whole algebra, and the general form of the
conserved charges is

\be
{\cal J}(x,p) = \hat{c} + \hat{a} \cdot p + \frac{1}{2}\, \hat{\og} \cdot M,
\label{6.9}
\ee

\nit
in which the co-efficients $(\hat{c}, \hat{a}, \hat{\og})$ are arbitrary
functions of the quad\-rat\-ic Cas\-im\-ir invariants $(p^{2}, M^{2})$.

\section{Motions of Spinning Points in Curved Space-Time}

\subsection{Spinning Space}

In this chapter we extend the spaces considered previously with additional
fermionic dimensions, parametrised by vectorial Grassmann co-ordinates
$\ps^{\mu}$. Following refs.\ct{BM}-\ct{BVH}, we take the extension in such a
way, that a supersymmetry is realised in these graded spaces; it acts on the
co-ordinates as

\be
\del x^{\mu} = -i \eps \ps^{\mu}, \hspace{4em}
                \del \ps^{\mu} = \eps \dot{x}^{\mu}.
\label{7.1}
\ee

\nit
Such graded spaces were called spinning spaces in \ct{RvH}. An action for the
extremal trajectories (`geodesics') of spinning space is

\be
S = \int_{1}^{2} d\tau\, \left( \frac{1}{2}\, g_{\mu\nu}(x) \dot{x}^{\mu}
     \dot{x}^{\nu} + \frac{i}{2}\, g_{\mu\nu}(x) \ps^{\mu}
     \frac{D \ps^{\nu}}{D\tau} \right),
\label{7.2}
\ee

\nit
where the covariant derivative of $\ps^{\mu}$ is defined by

\be
\frac{D \ps^{\mu}}{D\tau}\, =\, \dot{\ps}^{\mu}\, +\, \dot{x}^{\lb}
       \Gam_{\lb\nu}^{\:\:\:\mu} \ps^{\nu}.
\label{7.3}
\ee

\nit
Under a general variation of the the co-ordinates $(\del x^{\mu},
\del \ps^{\mu})$ the action changes by

\be
\ba{ll}
\del S = & \dsp{ \int_{1}^{2} d\tau\, \left\{ - \del x^{\mu}\, \left(
g_{\mu\nu}
         \frac{D^{2}x^{\nu}}{D \tau^{2}}\, +\, \frac{i}{2}\, \ps^{\kg}
\ps^{\lb}
         R_{\kg\lb\mu\nu}\, \dot{x}^{\nu} \right) \right. }\\
         &  \\
         & \dsp{\left. +\, i \Del \ps^{\mu} g_{\mu\nu}
\frac{D\ps^{\nu}}{D\tau}\,
           +\, \frac{d}{d\tau}\, \left( \del x^{\mu} p_{\mu} - \frac{i}{2}\,
           \del \ps^{\mu} g_{\mu\nu} \ps^{\nu} \right) \right\}. }
\ea
\label{7.4}
\ee

\nit
Here the canonical momentum is

\be
p_{\mu}\, =\, g_{\mu\nu} \dot{x}^{\nu}\, -\, \frac{i}{2}\, \Gam_{\mu\kg\lb}\,
          \ps^{\kg} \ps^{\lb},
\label{7.5}
\ee

\nit
whilst $R_{\kg\lb\mu\nu}$ is the Riemann curvature tensor. Moreover, we have
simplified the expression for $\del S$ by introducing a covariantized variation
of $\ps^{\mu}$:

\be
\Del \ps^{\mu}\, =\, \del \ps^{\mu}\, +\, \del x^{\lb}\,
                  \Gam_{\lb\nu}^{\:\:\:\mu}\, \ps^{\nu}.
\label{7.6}
\ee

\nit
The action $S$ is stationary under arbitrary variations $\del x^{\mu}$ and
$\del \ps^{\mu}$ vanishing at the end points if the following equations of
motion are satisfied

\bea
\dsp{ \frac{D^{2}x^{\mu}}{D\tau^{2}} } & = & \dsp{ - \frac{i}{2}\, \ps^{\kg}
      \ps^{\lb} R_{\kg\lb\:\:\nu}^{\:\:\:\:\mu}\, \dot{x}^{\nu}, }
\label{7.7a} \\
    &  & \nb \\
\dsp{ \frac{D\ps^{\mu}}{D\tau} } & = & 0.
\label{7.7b}
\eea

\nit
We briefly consider the physical interpretation of these equations. The
quantity

\be
S^{\mu\nu} = -i \ps^{\mu} \ps^{\nu}
\label{7.8}
\ee

\nit
can formally be regarded as the spin-polarisation tensor of the particle
\ct{BM}-\ct{BCL2},\ct{JW}, and correspondingly eqs.(\ref{7.7a},\ref{7.7b})
describe the classical motion of a Dirac particle. Eq.(\ref{7.7b}) implies that
the spin tensor is covariantly constant. Equation (\ref{7.7a}) then becomes

\be
\dsp{ \frac{D^{2}x^{\mu}}{D\tau^{2}} } =  \frac{1}{2}\, S^{\kg\lb}\,
R_{\kg\lb\:\:\nu}^{\:\:\:\:\mu}\, \dot{x}^{\nu}.
\label{7.8.1}
\ee

\nit
It implies, that there exist spin-dependent gravitational forces similar to the
electro-magnetic Lorentz force

\be
\ddot{x}^{\mu} = \frac{q}{m}\, F^{\mu}_{\:\:\nu}\, \dot{x}^{\nu},
\label{7.9}
\ee

\nit
with the spin-polarisation tensor replacing the scalar electric charge
\ct{JW,Kr} (here for unit mass). Such forces in principle allow the
determination of spin without any reference to the intrinsic electro-magnetic
dipole moments which are associated with it for charged particles.

\subsection{Symmetries and Generalised Killing Equations}

We now look for specific variations $\del x^{\mu}$ and $\Del \ps^{\mu}$ which
leave the action off-shell invariant modulo boundary terms. We take the
variations to be of the form

\be
\ba{l}
\dsp{ \del x^{\mu} = {\cal R}^{\mu}(x,\dot{x},\ps) = R^{(1)\mu}(x,\ps)\, +\,
    \sum_{n=1}^{\infty}\, \frac{1}{n!}\, \dot{x}^{\nu_{1}}...\dot{x}^{\nu_{n}}
    R_{\:\nu_{1}...\nu_{n}}^{(n+1)\:\mu}(x,\ps) } \\
    \\
\dsp{ \Del \ps^{\mu} = {\cal S}^{\mu}(x,\dot{x},\ps) = S^{(0)\mu}(x,\ps)\, +\,
    \sum_{n=1}^{\infty}\, \frac{1}{n!}\, \dot{x}^{\nu_{1}}...\dot{x}^{\nu_{n}}
    S_{\nu{1}...\nu_{n}}^{(n)\:\:\:\:\mu}(x,\ps) } \\
\ea
\label{8.1}
\ee

\nit
If the Lagrangian transforms into a total derivative

\be
\del S = \int_{1}^{2} d\tau\, \frac{d}{d\tau} \left( \del x^{\mu} p_{\mu} -
  \frac{i}{2}\, \del \ps^{\mu} g_{\mu\nu} \ps^{\nu} - {\cal J}(x,\dot{x}, \ps)
  \right),
\label{8.2}
\ee

\nit
it follows that

\be
\frac{d {\cal J}}{d \tau} = {\cal R}^{\mu} \left( g_{\mu\nu}
    \frac{D^{2}x^{\nu}}{D \tau^{2}} + \frac{i}{2}\, \ps^{\kg} \ps^{\lb}
    R_{\kg\lb\mu\nu} \dot{x}^{\nu} \right)\, +\, i\, {\cal S}^{\mu}
    g_{\mu\nu} \frac{D \ps^{\nu}}{D \tau}.
\label{8.3}
\ee

\nit
If the equations of motion are satisfied, the right-hand side vanishes and
${\cal J}$ is conserved. Again, this is Noether's theorem. Otherwise, expanding
${\cal J}(x,\dot{x},\ps)$ in terms of the four-velocity

\be
{\cal J}(x,\dot{x},\ps) = J^{(0)}(x,\ps)\, +\, \sum_{n=1}^{\infty}\,
   \frac{1}{n!}\, \dot{x}^{\mu_{1}}...\dot{x}^{\mu_{n}}
   J_{\mu_{1}...\mu_{n}}^{(n)}(x,\ps)
\label{8.4}
\ee

\nit
and comparing the left- and right-hand side of eq.(\ref{8.3}) with the Ansatz
(\ref{8.1}) for $\del x^{\mu}$ and $\Del \ps^{\mu}$, we find the following
identities

\be
J^{(n)}_{\mu_{1}...\mu_{n}}(x,\ps) = R^{(n)}_{\mu_{1}...\mu_{n}}(x,\ps),
\hspace{2em} n \geq 1;
\label{8.5}
\ee

\nit
and

\be
S^{(n)}_{\mu_{1}...\mu_{n}\nu}(x,\ps) =
                         i \dd{J^{(n)}_{\mu_{1}...\mu_{n}}}{\ps^{\nu}}(x,\ps),
\hspace{2em} n\geq 0.
\label{8.6}
\ee

\nit
Moreover, these quantities have to satisfy a generalisation of the Killing
equations of the form \ct{RvH}

\be
J^{(n)}_{\left( \mu_{1} ... \mu_{n};\mu_{n+1} \right)} +
   \dd{J^{(n)}_{\left( \mu_{1}...\mu_{n} \right.}}{\ps^{\sg}}\, \Gam_{\left.
   \mu_{n+1} \right) \kg}^{\hspace{2em}\:\:\:\sg}\, \ps^{\kg}\, =\,
\frac{i}{2}\,
   \ps^{\kg} \ps^{\lb} R_{\kg\lb\nu \left( \mu_{n+1} \right.}\,
   J^{(n+1)\:\:\nu}_{\left. \mu_{1}...\mu_{n} \right)}.
\label{8.7}
\ee

\nit
Writing as before $R^{(1)}_{\mu} = R_{\mu}$, $R^{(2)}_{\mu\nu} = K_{\mu\nu}$,
$R^{(3)}_{\mu\nu\lb} = L_{\mu\nu\lb}$, etc., and $J^{(0)} = B$, this reduces
for the lowest components to

\be
\ba{rcl}
\dsp{ B_{,\mu}\, +\, \dd{B}{\ps^{\sg}}\, \Gam_{\mu\kg}^{\:\:\:\:\sg}\,
    \ps^{\kg} } & = & \dsp{ \frac{i}{2}\, \ps^{\rg} \ps^{\sg}\,
R_{\rg\sg\kg\mu}
    \, R^{\kg}, } \\
  &  &  \\
\dsp{ R_{\left( \mu;\nu \right)}\, +\, \dd{R_{\left( \mu \right.}}{\ps^{\sg}}
    \, \Gam_{\left. \nu \right) \kg}^{\:\:\:\:\:\:\sg}\, \ps^{\kg} } & = &
\dsp{
    \frac{i}{2}\, \ps^{\rg} \ps^{\sg}\, R_{\rg\sg\kg\left( \mu \right.}\,
    K^{\:\:\:\:\:\kg}_{\left. \nu \right)}, } \\
  &  &  \\
\dsp{ K_{\left( \mu\nu;\lb \right)}\, +\, \dd{K_{\left( \mu\nu \right.}}{
    \ps^{\sg}}\, \Gam_{\left. \lb \right) \kg}^{\:\:\:\:\:\:\sg}\,
    \ps^{\kg} } & = & \dsp{ \frac{i}{2}\, \ps^{\rg} \ps^{\sg}\,
    R_{\rg\sg\kg\left( \mu \right.}\, L^{\hspace{1em}\:\:\kg}_{\left. \nu\lb
\right)}, }
    \hspace{4em} etc. \\
\ea
\label{8.8}
\ee

\nit
These equations have to hold independent of the equations of motion. The purely
bosonic ($\ps$-independent) parts of these equations reduce to those we found
for the scalar particle, eqs.(\ref{3.4a}-\ref{3.4c}). In particular, the
bosonic
terms in the Killing vectors $R^{\mu}$ define a Lie-algebra by taking
Lie-derivatives as in eq.(\ref{4.7a}). Furthermore we note that, contrary to
the bosonic case, the Killing scalar $B(x,\ps) = J^{(0)}(x,\ps)$ is not always
an irrelevant constant, because it can depend non-trivially on $x^{\mu}$ and
$\ps^{\mu}$, as follows from the first of eqs.(\ref{8.8}).

\subsection{Universal Solutions for Spinning Space}

In contrast to the scalar particle, the spinning particle admits several
conserved quantities of motion in a general curved space-time with metric
$g_{\mu\nu}(x)$ \ct{RvH}. Specifically, we can construct the following four
universal constants of motion: \nl

\nit
1. Like in the bosonic case $g_{\mu\nu}$ itself is a Killing tensor:

\be
K_{\mu\nu} = g_{\mu\nu},
\label{9.1}
\ee

\nit
with all other Killing vectors and tensors (bosonic as well as fermionic)
equal to zero. The corresponding constant of motion is the Hamiltonian

\be
H(x,P) = \frac{1}{2}\, g^{\mu\nu}(x) P_{\mu} P_{\nu},
\label{9.2}
\ee

\nit
where we have defined a covariant momentum

\be
P_{\mu} = p_{\mu}\, +\, \frac{i}{2}\, \Gam_{\mu\kg\lb}\, \ps^{\kg}
   \ps^{\lb}.
\label{9.3}
\ee

\nit
2. A second obvious solution is provided by the Grassmann-odd Killing vectors

\be
R^{\mu} = \ps^{\mu}, \hspace{2cm} T_{\mu}^{\nu} = i \del_{\mu}^{\nu}.
\label{9.4}
\ee

\nit
Again all other Killing vectors and tensors are taken to vanish. This solution
gives us the supercharge

\be
Q = P_{\mu} \ps^{\mu}.
\label{9.5}
\ee

\nit
3. In addition to ordinary supersymmetry, the spinning particle action has a
second non-linear supersymmetry, generated by Killing vectors

\be
\ba{lll}
R_{\mu} & = & \dsp{ \frac{-i^{[\frac{d}{2}]}}{(d-1)!}\, \sqrt{-g}\,
          \ve_{\mu\nu_{1}...\nu_{d-1}}\, \ps^{\nu_{1}}...\ps^{\nu_{d-1}}, } \\
  &  &  \\
T_{\mu\nu} & = & \dsp{ \frac{i^{[\frac{d-2}{2}]}}{(d-2)!}\, \sqrt{-g}\,
          \ve_{\mu\nu\nu_{1}...\nu_{d-2}}\, \ps^{\nu_{1}}...\ps^{\nu{d-2}}. }\\
\ea
\label{9.6}
\ee

\nit
Obviously, the Grassmann parities of $(R_{\mu}, T_{\mu\nu})$ depend on the
number of space-time dimensions. The corresponding constant of motion is the
dual supercharge, given by

\be
Q^{*} = \frac{-i^{[\frac{d}{2}]}}{(d-1)!}\, \sqrt{-g}\,
        \ve_{\mu_{1}...\mu_{d}}\, P^{\mu_{1}} \ps^{\mu_{2}}...\ps^{\mu_{d}}.
\label{9.7}
\ee

\nit
4. Finally, there exists a non-trivial Killing scalar

\be
\Gam_{*} \equiv J^{(0)} = - \frac{i^{[\frac{d}{2}]}}{d!}\, \sqrt{-g}\,
          \ve_{\mu_{1}...\mu_{d}}\, \ps^{\mu_{1}}...\ps^{\mu_{d}}.
\label{9.8}
\ee

\nit
This constant of motion acts as the Hodge star duality operator on $\ps^{\mu}$.
In quantum mechanics it becomes the $\gam^{d+1}$ element of the Dirac algebra.
For this reason $\Gam_{*}$ is refered to as the chiral charge.

{}From the fundamental Dirac brackets

\bea
\left\{ x^{\mu}, p_{\nu} \right\} &=& \del^{\mu}_{\nu}, \nonumber\\
\left\{ \ps^{\mu}, \ps^{\nu} \right\} &=& -i g^{\mu\nu}, \nonumber\\
\left\{ p_{\mu}, \ps^{\nu} \right\} &=& \textstyle{\frac{1}{2}}
g^{\kg \nu} g_{\kg \lb, \mu} \ps^{\lb}, \nonumber\\
\left\{ p_{\mu}, p_{\nu} \right\} &=& - \textstyle{\frac{i}{4}}
g^{\kg\lb} g_{\kg \rg, \mu} g_{\lb \sg, \nu} \ps^{\rg} \ps^{\sg}.
\label{9.9}
\eea

\nit
we now find the following non-trivial Dirac-brackets between these universal
constants of motion:

\be
\left\{ Q,Q \right\} = -2i\, H, \hspace{1.5cm}
\left\{ Q, \Gam_{*} \right\} = -i Q^{*}.
\label{9.10}
\ee

\nit
Observe, that $d = 2$ is an exceptional case: $Q^{*}$ is linear and acts as an
ordinary supersymmetry:

\be
\left\{ Q^{*}, Q^{*} \right\} = -2i\, H, \hspace{1.5cm}
\left\{ Q^{*}, \Gam_{*} \right\} = -i Q.
\label{9.11}
\ee

\nit
Hence in two dimensions the theory actually possesses an $N = 2$ supersymmetry.
For $d \neq 2$, the right-hand side of eqs.(\ref{9.11}) is to be replaced by
zero.

\section{Spinning Particles in Schwarzschild Space-Time}

\subsection{Conservation Laws in Schwarzschild Space-Time}\label{S3}

As an application of the generalised Killing equations for spinning space  we
discuss the motion of a spinning particle in a static and spherically symmetric
gravitational field\footnote{More details can be found in ref.\ct{RvH2}.}. The
field is described by the Schwarzschild metric

\be
ds^{2} = - \left( 1 - \frac{\ag}{r} \right) dt^{2} + \frac{1}{\left(1 -
         \frac{\ag}{r} \right)} dr^{2} + r^{2}\left( d\th^{2} + \sin^{2} \th
         d\vf^{2} \right),
\label{10.1}
\ee

\nit
where $\ag = 2 MG$, $M$ being the total mass of the spherically symmetric
object
in the centre of the field. It is well-known, that the Schwarzschild metric
possesses four Killing vector fields of the form

\be
D^{(\ag)} \equiv R^{(\ag) \mu}(x) \partial_{\mu}; \hspace{1.5cm} \ag = 0,...,3
\label{10.2}
\ee

\nit
where

\be
\ba{lll}
D^{(0)} & = & \dsp{ \dd{}{t}, }\\
 & & \\
D^{(1)} & = & \dsp{ -\sin \vf \dd{}{\th}\, +\, \cot \th \cos \vf \dd{}{\vf},
}\\
 & & \\
D^{(2)} & = & \dsp{ \cos \vf \dd{}{\th}\, -\, \cot \th \sin \vf \dd{}{\vf}, }\\
 & & \\
D^{(3)} & = & \dsp{ \dd{}{\vf}. }\\
\ea
\label{10.3}
\ee

\nit
These Killing vector fields express the time-translation invariance and the
rotation symmetry of the gravitational field. They generate the corresponding
Lie algebra {\large o}(1,1) $\times$ {\large so}(3):

\be
\ba{lll}
\left[ D^{(i)}, D^{(j)} \right] & = & - \ve^{ijk} D^{(k)}, \\
  &  &  \\
\left[ D^{(0)}, D^{(i)} \right] & = & 0. \\
\ea
\label{10.4}
\ee

\nit
The first generalised Killing equation (\ref{8.8}) shows, that with each
Killing
vector $R^{(\ag)}_{\mu}$ there is an associated Killing scalar $B^{(\ag)}$.
These Killing scalars are necessary to obtain the constants of motion

\be
J^{(\ag)} = B^{(\ag)} + m \dot{x}^{\mu} R^{(\ag)}_{\mu}.
\label{10.5}
\ee

\nit
The Killing scalars have a natural interpretation: the constants of motion
represent the {\em total} angular momentum, which is the sum of the orbital
and the spin angular momentum. In general, orbital angular momentum is no
longer
separately conserved; therefore the Killing vector itself does {\em not} give a
conserved quantity of motion. The contribution of spin is contained in the
Killing scalars, and has to be added.

Inserting the expression for the connections and the Riemann curvature
components of the Schwarzschild space-time in eq.(\ref{8.8}), we obtain for the
Killing scalars

\be
\ba{lll}
B^{(0)} & = & \dsp{ \frac{-i \ag}{2r^{2}}\, \ps^{t} \ps^{r}, }\\
  & & \\
B^{(1)} & = & \dsp{ i r \sin \vf \ps^{r} \ps^{\th}\, +\, i r \sin \th \cos \th
              \cos \vf \ps^{r} \ps^{\vf}\, -\, i r^{2} \sin^{2} \th \cos \vf
              \ps^{\th} \ps^{\vf}, }\\
  & & \\
B^{(2)} & = & \dsp{ -i r \cos \vf \ps^{r} \ps^{\th}\, +\, i r \sin \th \cos \th
              \sin \vf \ps^{r} \ps^{\vf}\, -\, i r^{2} \sin^{2} \th \sin \vf
              \ps^{\th} \ps^{\vf}, }\\
  & & \\
B^{(3)} & = & \dsp{ -i r \sin^{2} \th \ps^{r} \ps^{\vf}\, -\, i r^{2} \sin \th
              \cos \th \ps^{\th} \ps^{\vf}. }\\
\ea
\label{10.6}
\ee

\nit
Upon subtitution in $J^{(\ag)}$, eq.(\ref{10.5}), and using the spin-tensor
notation introduced in eq.(\ref{7.8}), one finds

\be
\ba{lll}
J^{(0)} & \equiv & E = \dsp{ m \left( 1 - \frac{\ag}{r} \right)
                   \frac{dt}{d\tau}\, -\, \frac{\ag}{2 r^{2}}\, S^{rt}, }\\
  & & \\
J^{(1)} & = & \dsp{ - r \sin \vf \left( mr \frac{d\th}{d\tau} + S^{r\th}
\right)
        \, -\, \cos \vf \left( \cot \th J^{(3)} - r^{2} S^{\th\vf} \right), }\\
  & & \\
J^{(2)} & = & \dsp{ r \cos \vf \left( mr \frac{d\th}{d\tau} + S^{r\th}
\right)\,
        -\, \sin \vf \left( \cot \th J^{(3)} - r^{2} S^{\th\vf} \right), }\\
  & & \\
J^{(3)} & = & \dsp{ r \sin^{2} \th \left( mr \frac{d\vf}{d\tau} + S^{r\vf}
        \right)\, +\, r^{2} \sin \th \cos \th S^{\th\vf}. }\\
\ea
\label{10.7}
\ee

\nit
In addition to these constants of motion, the universal conserved charges
such as the world-line hamiltonian and supercharge also provide information
about the allowed orbits of the particle. Specifically, we consider motions for
which

\be
H = \frac{-m^{2} c^{2}}{2}, \hspace{2cm} Q = 0.
\label{10.8}
\ee

\nit
The first equation implies geodesic motion (equality of proper time with the
geodesic interval):

\be
d\tau^{2} = - ds^{2},
\label{10.9}
\ee

\nit
cf.\ eqs.(\ref{1.2}),(\ref{10.1}). The second equation expresses the fact that
spin represents only three independent degrees of freedom. Indeed, we can now
solve for $\ps^{t}$ in terms of the spatial components $\ps^{i}$:

\be
\left( 1 - \frac{\ag}{r} \right)\, \frac{dt}{d\tau}\, \ps^{t}\, = \,
  \frac{1}{\left( 1 - \frac{\ag}{r} \right)}\, \frac{dr}{d\tau}\, \ps^{r}\, +\,
  r^{2} \left( \frac{d\th}{d\tau} \ps^{\th} + \sin^{2} \th \frac{d\vf}{d\tau}\,
  \ps^{\vf} \right).
\label{10.10}
\ee

\nit
As a result, the (classical) chiral charge $\Gam_{*}$ vanishes as well.

{}From eqs.(\ref{10.7}) one can derive a useful identity

\be
r^{2} \sin \th\, S^{\th \vf}\, =\, J^{(1)} \sin \th \cos \vf\, +\, J^{(2)} \sin
      \th \sin \vf\, +\, J^{(3)} \cos \th.
\label{10.11}
\ee

\nit
In physical terms, this identity simply states that there is no orbital angular
momentum in the radial direction.

Combining eqs.(\ref{10.7}),(\ref{10.8}) one obtains a complete set of first
integrals of motion, expressing the velocities as functions of the
co-ordinates,
the spatial spin components and the constants of motion:

\be
\ba{lll}
\dsp{ \frac{dt}{d\tau} } & = & \dsp{ \frac{1}{\left(1 - \frac{\ag}{r} \right)}
      \left( \frac{E}{m} + \frac{\ag}{2mr^{2}}\, S^{rt} \right), }\\
  &  &  \\
\dsp{ \frac{dr}{d\tau} } & = & \dsp{ \left[ \left(1 - \frac{\ag}{r} \right)^{2}
      \left( \frac{dt}{d\tau} \right)^{2} - 1 + \frac{\ag}{r} \right. }\\
  &  &  \\
  &  & \dsp{ \left. \:\: - r^{2} \left( 1 - \frac{\ag}{r} \right) \left\{
       \left( \frac{d\th}{d\tau} \right)^{2} + \sin^{2} \th \left(
       \frac{d\vf}{d\tau} \right)^{2} \right\} \right]^{1/2}, }\\
  &  &  \\
\dsp{ \frac{d\th}{d\tau} } & = & \dsp{ \frac{1}{mr^{2}}\, \left( - J^{(1)}
      \sin \vf + J^{(2)} \cos \vf - r S^{r\th} \right), }\\
  &  &  \\
\dsp{ \frac{d\vf}{d\tau} } & = & \dsp{ \frac{1}{mr^{2} \sin^{2} \th}\, J^{(3)}
-
      \frac{1}{mr}\, S^{r\vf} - \frac{1}{m} \cot \th\, S^{\th\vf}, }\\
\ea
\label{10.12}
\ee

\nit
in which

\be
S^{rt} = \frac{mr^{2}}{E}\, \left( \frac{d\th}{d\tau}\, S^{r\th} + \sin^{2} \th
         \frac{d\vf}{d\tau}\, S^{r\vf} \right).
\label{10.13}
\ee

\nit
Finally, the rate of change of the spins is determined by

\be
\ba{lll}
\dsp{ \frac{d\ps^{r}}{d\tau} } & = & \dsp{ r \left( 1 - \frac{3\ag}{2r} \right)
      \left( \frac{d\th}{d\tau}\, \ps^{\th} + \sin^{2} \th \frac{d\vf}{d\tau}\,
      \ps^{\vf} \right), }\\
  &  &  \\
\dsp{ \frac{d\ps^{\th}}{d\tau} } & = & \dsp{ - \frac{1}{r}\, \left( \frac{dr}{d
      \tau}\, \ps^{\th} + \frac{d\th}{d\tau}\, \ps^{r} \right) + \sin \th
      \cos \th \frac{d\vf}{d\tau}\, \ps^{\vf}, }\\
  &  &  \\
\dsp{ \frac{d\ps^{\vf}}{d\tau} } & = & \dsp{ - \left( \frac{1}{r} \frac{dr}{d
      \tau}\, + \cot \th \frac{d\th}{d\tau} \right)\, \ps^{\vf} - \frac{1}{r}
      \frac{d\vf}{d\tau}\, \ps^{r} - \cot \th \frac{d\vf}{d\tau}\, \ps^{\th}.
      }\\
\ea
\label{10.14}
\ee

\nit
Eqs.(\ref{10.12}-\ref{10.14}) can be integrated to give the full solution of
the
equations of motion for all co-ordinates and spins.

\subsection{Special Solutions}

As an application of the results obtained in sect.(\ref{S3}) we study the
special case of motion in a plane, for which we choose $\th = \pi/2$. Unlike
for
scalar point particles, this is not the generic case, because in general
orbital angular momentum is not conserved separately.

For spinning particles, motion in a plane occurs in two kind of situations. The
first possibility is radial motion, for which $\dot{\vf} = 0$. In this case
there is no orbital angular momentum and spin is conserved independently.
Clearly, either the particle escapes to infinity or hits the centre of the
potential after a finite time.

The second possibility concerns motion for which $\dot{\vf} \neq 0$. In this
case orbital and spin angular momentum decouple if they are parallel. Hence we
impose the conditions

\be
S^{\th\vf} = 0, \hspace{2cm} S^{r\th} = 0,
\label{11.1}
\ee

\nit
from which it now follows that $J^{(1,2)}$ vanish. In this case we have two
constants of motion:

\be
L = mr^{2} \dot{\vf}, \hspace{2cm} \Sg \equiv J^{(3)} - L = r S^{r\vf}.
\label{11.2}
\ee

\nit
{}From the first of eqs.(\ref{10.12}) we now find a formula for the
gravitational
redshift given by

\be
dt = \frac{d\tau}{\left(1 - \frac{\ag}{r}\right)}\, \left( \frac{E}{m} +
     \frac{\ag}{2mEr^{3}}\, L \Sg \right).
\label{11.3}
\ee

\nit
The first term corresponds to the usual time dilation in a gravitational field
for spinless particles. In this case, there is an additional contribution from
spin-orbit coupling. This shows, that time dilation is not a purely geometrical
effect, but also has a dynamical component \ct{JW}.

{}From the second of eqs.(\ref{10.12}) and the first eq.(\ref{11.2}) we obtain
the equation for the orbit of the particle:

\be
\frac{1}{r} \frac{dr}{d\vf}\, =\, \sqrt{ \dsp{ \frac{\left( E^{2} - m^{2}
     \right)}{L^{2}} r^{2} - 1 + \frac{m\ag}{L} \left( \frac{mr}{L} +
     \frac{J^{(3)}}{mr} \right) } }.
\label{11.4}
\ee

\nit
In terms of dimensionless quantities

\be
\ba{rclrcl}
\eps & = & \dsp{ \frac{E}{m}, }    &   x  & = & \dsp{ \frac{r}{\ag}, } \\
  &  &  &  &  &  \\
\ell & = & \dsp{ \frac{L}{m\ag}, } & \Del & = & \dsp{ \frac{\Sg}{L}, } \\
\ea
\label{11.5}
\ee

\nit
we find for the stationary points $x_{m}$ of the orbit, as defined by the
vanishing of the right-hand side of eq.(\ref{11.4}):

\be
\eps^{2}\, =\, 1\, -\, \frac{1}{x_{m}}\, +\, \frac{\ell^{2}}{x_{m}^{2}}\,
           -\, \frac{\ell^{2}}{x_{m}^{3}}\, \left( 1 + \Del \right).
\label{11.6}
\ee

\nit
Of course, all calculations presented here are rather formal, as $\Del$ is not
a
pure number. However, the equations might be applicable to realistic physics
situations if it is allowed to replace $\Del$ in certain limiting cases by a
real
number. Since the pseudo-classical equations acquire physical meaning when
averaged over in functional integrals \ct{BM,BCL3} (i.e.\ in the path integral
of the quantum Dirac particle), such a limit might arise in the semi-classical
regime of the quantum theory, as implied by the correspondence principle. In
the
following we assume that such a numerical value of $\Del$ has been obtained
and leads to valid results, at least in expansions to first order in $\Del$
(where the fact that $\Del^{2} = 0$ plays no role).

For $\eps \geq 1$ we have open orbits for which there is at most one point of
closest approach, the perihelion. If for fixed $\ell$ the energy exceeds a
critical value, the particle can cross into the central region of the potential
$(x < 1)$. The critical value is given by

\be
\eps_{crit}^{2} \, =\, \eps^{2}(x^{(-)}_{m}),
\label{11.7}
\ee

\nit
where

\be
x^{(-)}_{m} = \ell^{2} \left( 1 - \sqrt{1 - \frac{3 (1 + \Del)}{\ell^{2}} }
\right).
\label{11.8}
\ee

\nit
For $\eps < 1$ there are bound states corresponding to quasi-periodic orbits,
which have both a perihelion and an aphelion. The stationary points of the
orbit are again determined by\footnote{ For the special case of circular
motion,
this equation determines the radius of the orbit, which now is of course
independent of $\vf$.} eq.(\ref{11.6}). Solutions of this equation exist only
for

\be
\ell^{2} \geq 3 \left( 1 + \Del \right).
\label{11.9}
\ee

\nit
In particular, for circular orbits there exists a minimal radius given by

\be
x_{m} = \ell^{2} = 3 \left( 1 + \Del \right).
\label{11.10}
\ee

\nit
For this critical orbit the energy is to first order in $\Del$:

\be
\eps^{2}_{crit} = \frac{1}{9}\, \left (8 + \Del \right) ,
\label{11.11}
\ee

\nit
whilst the time-dilation factor in the critical orbit is given by

\be
\left( \frac{dt}{d\tau} \right)_{crit}\, =\, \sqrt{2} \left( 1- \frac{3}{8}
       \Del \right).
\label{11.12}
\ee

\nit
For non-circular motion one finds that the perihelion of the orbit precesses
as for a spinless particle, but at a different, spin dependent rate. In the
weak-coupling limit (slow precession of the perihelion), we obtain for the
precession angle after one period

\be
\Del \vf = \frac{3 \pi \ag}{k}\, \left( 1 + \Del \right),
\label{11.13}
\ee

\nit
where $k$ is the {\em semilatus rectum\/} of the elliptical orbit, and we have
neglected terms of $O(k^{-2})$. Hence in this approximation spin-dependent
effects disappear for $\Del = O(k^{-p})$ whenever $p \geq 1$. Note, that
spin-dependent gravitational effects can be larger or smaller than for a
spinless particle depending on the sign of $\Del$, i.e.\ the relative
orientation of $L$ and $\Sg$. This we interpret as a classical analogue of fine
splitting.

As a final remark, we observe that even if an {\em a priori\/} numerical value
for $\Del$ cannot be assigned, its appearance in various places like in
eqs.(\ref{11.11}-\ref{11.13}) still allows the pseudo-classical theory to make
quantitative predictions by comparing different physical processes in the
regime
where the semi-classical limit applies. For a discussion of some related issues
in the case of spinning particles in electro-magnetic fields we also refer to
\ct{JW}.

\end{document}